\documentclass{ws-procs9x6}

\begin{document}

\title{
  \vspace*{-2cm}
  \hfill{\normalsize\vbox{%
    \hbox{\rm\small DPNU-03-06}
  }}\\
  \vspace{0.2cm}
  Vector Manifestation of Chiral Symmetry~\footnote{%
  \uppercase{T}alk given at 2002 \uppercase{I}nternational 
  \uppercase{W}orkshop on
  ``\uppercase{S}trong \uppercase{C}oupling \uppercase{G}auge 
  \uppercase{T}heories and \uppercase{E}ffective \uppercase{F}ield 
  \uppercase{T}heories''
  (\uppercase{SCGT}02), \uppercase{D}ecember 10-13, 2002.
  \newline
  \uppercase{T}his talk is based on thw works done in 
  \uppercase{R}efs.~1, 2 and 3.
}
}

\author{M\lowercase{asayasu} H\lowercase{arada}~\footnote{%
  e-mail: harada@eken.phys.nagoya-u.ac.jp
}
}

\address{Department of Physics, Nagoya University, Nagoya 464-8602,
  Japan}

\maketitle

\abstracts{
In this talk I summarize main features of the vector
manifestation (VM) which was recently proposed
as a novel manifestation of the Wigner realization of 
chiral symmetry in which the symmetry is restored at the critical
point by the massless degenerate pion (and its flavor partners) and
the $\rho$ meson (and its flavor partners) as the chiral partner.
I show how the VM is formulated in 
the effective field theory of QCD based on the
hidden local symmetry and 
realized in the large flavor QCD as well as
in hot and/or dense QCD.
}

\section{Introduction}

\nocite{HY:VM,HS:VMT,HKR}
Spontaneous chiral symmetry breaking is one of the 
most important properties of QCD in low energy region.
When one increases the number of massless quarks (large $N_f$
QCD) or considers QCD in hot and/or dense matter (hot and/or dense
QCD), this chiral symmetry is expected to be
restored~\cite{rest:Nf,rest:HD,Brown-Rho:96,Brown-Rho:01b}.

Recently in Ref.~\refcite{HY:VM},
{\it the vector manifestation (VM)}
is proposed as a new pattern
of the Wigner realization of chiral
symmetry, in which 
the chiral symmetry is restored at the critical point by 
{\it the massless degenerate pion (and its flavor
partners) and the $\rho$ meson (and its flavor partners) as
the chiral partner},
in sharp contrast to the traditional manifestation \`a la linear
sigma model where the symmetry is restored by the degenerate pion and
the scalar meson.
The formulation of the VM was done in the framework of the
effective field theory of QCD
based on the hidden local symmetry (HLS)~\cite{BKUYY-BKY:88}
where,
{\it thanks to the gauge invariance}, one can perform a
systematic loop expansion including the vector mesons in addition to
the pseudoscalar mesons.~\cite{Georgi,HY:PLB,Tanabashi,HY:WM,HY:PRep}
Essential roles to realize the VM are played
by the Wilsonian matching~\cite{HY:WM} between the HLS and the 
underlying QCD which determines the bare parameters of the HLS
Lagrangian, and the Wilsonian renormalization group equations
(RGEs) for the HLS 
parameters~\cite{HY:conformal} which include effects of
quadratic divergences.

The Wilsonian matching was applied to the analysis in hot QCD
in Ref.~\refcite{HS:VMT}, and it was stressed that 
the bare parameters of the HLS Lagrangian have the
{\it intrinsic temperature dependence}.
Then, it was shown that the intrinsic temperature dependence
plays an essential role for the VM to be realized at the critical
temperature in hot QCD.
In Ref.~\refcite{HKR}, 
following the picture shown in Ref.~\refcite{Brown-Rho:01b},
the quasiquark degree of freedom is added to the HLS Lagrangian
for analyzing dense QCD near the critical point.
Then, it was shown that the {\it intrinsic density dependences}
of the bare parameters of the HLS Lagrangian lead to the
VM in dense QCD at the critical chemical potential
in accord with the Brown-Rho scaling~\cite{Brown-Rho:96}.

In this write-up,
I first summarize main features of the VM comparing it with the
conventional picture based on the linear sigma model.
Then, I will show how the VM is formulated in the HLS
and
realized in the large $N_f$ QCD as well as
in hot and/or dense QCD.

This write-up is organized as follows.
In section~\ref{sec:VM}, I briefly explain some features of the
VM, and discuss the difference between
the VM and the conventional picture based on the linear sigma model.
In section~\ref{sec:EFT}, I briefly review the HLS model,
and show the Wilsonian RGEs for the HLS parameters.
Section~\ref{sec:WM} is devoted to summarize the 
Wilsonian matching.
In sections~\ref{sec:VMNf}, \ref{sec:VMT} and \ref{sec:VMm},
I show how the VM is realized in 
the large $N_f$ QCD as well as
in hot and/or dense QCD.
Finally, I give a brief summary in section~\ref{sec:sum}.

\section{Vector Manifestation}
\label{sec:VM}

In this section I briefly explain some features of the
vector manifestation (VM).
The VM was first proposed in
Ref.~\refcite{HY:VM} as a novel manifestation of Wigner 
realization of
chiral symmetry where the vector meson $\rho$ becomes massless at the
chiral phase transition point. 
Accordingly, the (longitudinal) $\rho$ becomes the chiral partner of
the Nambu-Goldstone boson $\pi$.

The VM is characterized by
\begin{equation}
F_\pi^2 \rightarrow 0 \ , \quad
m_\rho^2 \rightarrow m_\pi^2 = 0 \ , \quad
F_\rho^2 / F_\pi^2 \rightarrow 1 \ ,
\label{VM def}
\end{equation}
where $F_\rho$ is the decay constant of 
(longitudinal) $\rho$ at $\rho$ on-shell.
This is completely different from the conventional picture based
on the linear sigma model (I call this GL manifestation after the
effective theory of Ginzburg--Landau or Gell-Mann--Levy.)
where the scalar meson becomes massless
degenerate with $\pi$ as the chiral partner.
Here, I
discuss the difference between the VM and the
GL manifestation in terms of the chiral representation of the mesons
by extending the analyses done in
Refs.~\refcite{Gilman-Harari} and \refcite{Weinberg:69}
for two flavor QCD.
Since we are approaching the chiral restoration point only {\it from
the broken phase} where the chiral symmetry is realized only
nonlinearly, 
it does not make sense to discuss the chiral representation of such a 
spontaneously broken symmetry. 
Then we need a tool to formulate the {\it linear representation}
 of the chiral algebra
even {\it in the broken phase}, 
namely the classification algebra valid even in the broken phase, in such a way that it smoothly moves over to the 
original chiral algebra as we go over to the symmetric phase.

Following Ref.~\refcite{Weinberg:69},
I define the
axialvector coupling matrix $X_a(\lambda)$ 
(an analog of the $g_A$ for
the nucleon matrix)
by giving the matrix elements at zero invariant momentum transfer 
of
the axialvector current between states with collinear momenta
as
\begin{equation}
\langle \vec{q} \,\lambda^\prime \, \beta \vert
J_{5a}^+ (0)
\vert \vec{p} \,\lambda \, \alpha \rangle
=
2 
p^+\, \delta_{\lambda\lambda^\prime}
\left[ X_a(\lambda) \right]_{\beta\alpha}
\ ,
\label{axial coupling}
\end{equation}
where $J_{5a}^+ = (J_{5a}^{0} + J_{5a}^{3} )/\sqrt{2}$,
and $\alpha$ and $\beta$ are one-particle states
with collinear momentum $\vec{p}\equiv (p^+,p^1,p^2)$ and $\vec{q}
\equiv (q^+,q^1,q^2)$ such that $p^+=q^+$, 
$\lambda$ and $\lambda^\prime$ are their helicities.
It was stressed~\cite{Weinberg:69} that the definition
of the axialvector couplings in Eq.~(\ref{axial coupling}) can be 
used for particles of arbitrary spin, and in arbitrary collinear
reference frames, including both the frames in which $|\alpha\rangle$
is at 
rest and in which it moves with infinite momentum:
The matrix $X_a(\lambda)$ is independent of the reference frame. 
Note that the $X_a(\lambda)$ matrix does not contain the $\pi$ pole
term 
which would behave as $(p^+ -q^+)/[(p-q)^2 -m_\pi^2]$ and hence be
zero for 
kinematical reason, $p^+ = q^+$, even in the chiral limit of 
$m_\pi^2 \rightarrow 0$.

As was done for $N_f=2$ in Ref.~\refcite{Weinberg:69},
considering
the forward scattering process
$\pi_a + \alpha(\lambda) \rightarrow \pi_b + \beta(\lambda^\prime)$
and 
requiring the cancellation of the terms in the $t$-channel,
we obtain
\begin{equation}
\left[ \, X_a(\lambda)\,,\, X_b(\lambda) \,\right]
= i f_{abc} T_c
\ ,
\label{algebra}
\end{equation}
where
$T_c$ is the generator of $\mbox{SU}(N_f)_{\rm V}$ and 
$f_{abc}$ is the structure constant. 
This is nothing but the algebraization
of the Adler-Weisberger sum rule~\cite{Adler-Weisberger} 
and the basis of the good-old-days classification of the hadrons by
the chiral algebra~\cite{Gilman-Harari,Weinberg:69}
or the ``mended  symmetry.''~\cite{Weinberg:90}
It should be noticed that Eq.~(\ref{algebra}) tells us that the
one-particle states of any given helicity must be assembled into
representations of chiral 
$\mbox{SU}(N_f)_{\rm L}\times\mbox{SU}(N_f)_{\rm R}$.
Furthermore, since Eq.~(\ref{algebra}) does not give any relations 
among the states with different helicities, those states can 
generally belong to the different representations even though
they form a single particle such as the longitudinal $\rho$
($\lambda=0$) and the transverse $\rho$ ($\lambda=\pm1$).
Thus, the notion of the chiral partners can be considered
separately for each helicity.

Let me first consider the zero helicity ($\lambda=0$) states
and saturate the algebraic relation in Eq.~(\ref{algebra}) by low
lying mesons; 
the $\pi$, the (longitudinal) $\rho$, the (longitudinal) axialvector
meson denoted by $A_1$ ($a_1$ meson and its flavor partners)
and the scalar meson denoted by $S$, and so on.
The $\pi$ and the longitudinal $A_1$ 
are admixture of $(8\,,\,1) \oplus(1\,,\,8)$ and 
$(3\,,\,3^*)\oplus(3^*\,,\,3)$, since the symmetry is spontaneously
broken~\cite{Weinberg:69,Gilman-Harari}:
\begin{eqnarray}
\vert \pi\rangle &=&
\vert (3\,,\,3^*)\oplus (3^*\,,\,3) \rangle \sin\psi
+
\vert(8\,,\,1)\oplus (1\,,\,8)\rangle  \cos\psi
\ ,
\nonumber
\\
\vert A_1(\lambda=0)\rangle &=&
\vert (3\,,\,3^*)\oplus (3^*\,,\,3) \rangle \cos\psi 
- \vert(8\,,\,1)\oplus (1\,,\,8)\rangle  \sin\psi
\ ,
\label{mix pi A}
\end{eqnarray}
where the experimental value of the mixing angle $\psi$ is 
given by approximately 
$\psi=\pi/4$~\cite{Weinberg:69,Gilman-Harari}.  
On the other hand, the longitudinal $\rho$
belongs to pure $(8\,,\,1)\oplus (1\,,\,8)$
and the scalar meson to 
pure $(3\,,\,3^*)\oplus (3^*\,,\,3)$:
\begin{eqnarray}
\vert \rho(\lambda=0)\rangle &=&
\vert(8\,,\,1)\oplus (1\,,\,8)\rangle  
\ ,
\nonumber
\\
\vert S\rangle &=&
\vert (3\,,\,3^*)\oplus (3^*\,,\,3) \rangle 
\ .
\label{rhoS}
\end{eqnarray}

When the chiral symmetry is restored at the
phase transition point, the axialvector coupling matrix
commutes with the Hamiltonian matrix, and thus the 
chiral representations coincide with the mass eigenstates:
The representation mixing is dissolved.
{}From Eq.~(\ref{mix pi A}) one can easily see~\cite{HY:VM}
that
there are two ways to express the representations in the
Wigner phase of the chiral symmetry:
The conventional GL manifestation
corresponds to 
the limit $\psi \rightarrow \pi/2$ in which
$\pi$ is in the representation
of pure $(3\,,\,3^*)\oplus(3^*\,,\,3)$ 
[$(N_f\,,\,N_f^\ast)\oplus(N_f^\ast\,,\,N_f)$ of
$\mbox{SU}(N_f)_{\rm L} \times\mbox{SU}(N_f)_{\rm R}$ in 
large $N_f$ QCD]
together with the scalar meson, 
both being the chiral partners:
\begin{eqnarray}
\mbox{(GL)}
\qquad
\left\{
\begin{array}{rcl}
\vert \pi\rangle\,, \vert S\rangle
 &\rightarrow& 
\vert  (N_f\,,\,N_f^\ast)\oplus(N_f^\ast\,,\,N_f)\rangle\ ,
\\
\vert \rho (\lambda=0) \rangle \,,
\vert A_1(\lambda=0)\rangle  &\rightarrow&
\vert(N_f^2-1\,,\,1) \oplus (1\,,\,N_f^2-1)\rangle\ .
\end{array}\right.
\end{eqnarray}
On the other hand, the VM corresponds 
to the limit $\psi\rightarrow 0$ in which the $A_1$ 
goes to a pure 
$(3\,,\,3^*)\oplus (3^*\,,\,3)$
[$(N_f\,,\,N_f^\ast)\oplus(N_f^\ast\,,\,N_f)$], now degenerate with
the scalar meson in the same representation, 
but not with $\rho$ in 
$(8\,,\,1)\oplus (1\,,\,8)$
[$(N_f^2-1\,,\,1) \oplus (1\,,\,N_f^2-1)$]:
\begin{eqnarray}
\mbox{(VM)}
\qquad
\left\{
\begin{array}{rcl}
\vert \pi\rangle\,, \vert \rho (\lambda=0) \rangle
 &\rightarrow& 
\vert(N_f^2-1\,,\,1) \oplus (1\,,\,N_f^2-1)\rangle\ ,
\\
\vert A_1(\lambda=0)\rangle\,, \vert S\rangle  &\rightarrow&
\vert  (N_f\,,\,N_f^\ast)\oplus(N_f^\ast\,,\,N_f)\rangle\ .
\end{array}\right.
\end{eqnarray}
Namely, the
degenerate massless $\pi$ and (longitudinal) $\rho$ at the 
phase transition point are
the chiral partners in the
representation of $(8\,,\,1)\oplus (1\,,\,8)$
[$(N_f^2-1\,,\,1) \oplus (1\,,\,N_f^2-1)$].\footnote{
  It should be stressed 
  that the VM is realized only as a limit approaching
  the critical point from the broken phase 
  but not exactly on the critical point where the light spectrum
  including the $\pi$ and the $\rho$ would dissappear altogether.
}

Next, we consider the helicity $\lambda=\pm1$. 
As we stressed above,
the transverse $\rho$
can belong to the representation different from the one
for the longitudinal $\rho$ ($\lambda=0$) and thus can have the
different chiral partners.
According to the analysis in Ref.~\refcite{Gilman-Harari},
the transverse components of $\rho$ ($\lambda=\pm1$)
in the broken phase
belong to almost pure
$(3^*\,,\,3)$ ($\lambda=+1$) and $(3\,,\,3^*)$ ($\lambda=-1$)
with tiny mixing with
$(8\,,\,1)\oplus(1\,,\,8)$.
Then, it is natural to consider in VM that
they become pure $(N_f\,,\,N_f^\ast)$ and 
$(N_f^\ast\,,\,N_f)$
in the limit approaching the chiral restoration point:
\begin{eqnarray}
\vert \rho(\lambda=+1)\rangle \rightarrow 
  \vert (N_f^*,N_f)\rangle\ ,\quad
\vert \rho(\lambda=-1)\rangle \rightarrow 
  \vert (N_f,N_f^*)\rangle \ .
\end{eqnarray}
As a result,
the chiral partners of the transverse components of $\rho$ 
in the VM
will be  themselves. Near the critical point the longitudinal $\rho$
becomes 
almost $\sigma$, namely the would-be NG boson $\sigma$ almost 
becomes a 
true NG boson and hence a different particle than the transverse
$\rho$.

\section{Effective Field Theory}
\label{sec:EFT}

In this section I show the effective field theory (EFT)
in which
the vector manifestation is formulated.
I should note that,
as is stressed in Ref.~\refcite{HY:PRep},
the VM can be formulated only as a limit by approaching it from the
broken phase of chiral symmetry.
Then, for the formulation of the VM,
I need an EFT
including $\rho$ and $\pi$ 
in the broken phase which is not
necessarily applicable in the symmetric phase.
One of such EFTs is the model based on the 
hidden local symmetry (HLS)~\cite{BKUYY-BKY:88}
which includes $\rho$ as the gauge boson of the HLS in addition
to $\pi$ as the NG boson associated with the
chiral symmetry breaking in a manner fully consistent with
the chiral symmetry of QCD.
It should be noticed that,
in the HLS,
thanks to the gauge invariance
one can perform the
systematic chiral perturbation with including $\rho$
in addition to $\pi$.~\cite{Georgi,HY:PLB,Tanabashi,HY:WM,HY:PRep}
In subsection~\ref{ssec:HLS}, I will explain the model based
on the HLS, and then summarize the 
renormalization group equations (RGEs) for 
the parameters of the HLS Lagrangian in subsection~\ref{ssec:RGE}.

\subsection{Hidden Local Symmetry}
\label{ssec:HLS}

Let me describe the HLS model
based on the
$G_{\rm global} \times H_{\rm local}$ symmetry, where
$G = \mbox{SU($N_f$)}_{\rm L} \times 
\mbox{SU($N_f$)}_{\rm R}$  is the 
global chiral symmetry and 
$H = \mbox{SU($N_f$)}_{\rm V}$ is the HLS.
The basic quantities 
are the gauge boson 
$\rho_\mu$
and two 
variables 
\begin{eqnarray}
&&
\xi_{\rm L,R} = e^{i\sigma/F_\sigma} e^{\mp i\pi/F_\pi}
\ ,
\end{eqnarray}
where $\pi$
denotes the pseudoscalar NG boson
and $\sigma$
the NG boson absorbed into $\rho_\mu$ 
(longitudinal $\rho$).
$F_\pi$ and $F_\sigma$ are relevant decay constants, and
the parameter $a$ is defined as
$a \equiv F_\sigma^2/F_\pi^2$.
The transformation properties of $\xi_{\rm L,R}$ are given by
\begin{eqnarray}
&&
\xi_{\rm L,R}(x) \rightarrow \xi_{\rm L,R}^{\prime}(x) =
h(x) \xi_{\rm L,R}(x) g^{\dag}_{\rm L,R}
\ ,
\end{eqnarray}
where $h(x) \in H_{\rm local}$ and 
$g_{\rm L,R} \in G_{\rm global}$.
The covariant derivatives of $\xi_{\rm L,R}$ are defined by
\begin{eqnarray}
&&
D_\mu \xi_{\rm L} =
\partial_\mu \xi_{\rm L} - i g \rho_\mu \xi_{\rm L}
+ i \xi_{\rm L} {\mathcal L}_\mu
\ ,
\nonumber\\
&&
D_\mu \xi_{\rm R} =
\partial_\mu \xi_{\rm R} - i g \rho_\mu \xi_{\rm R}
+ i \xi_{\rm R} {\mathcal R}_\mu
\ ,
\label{covder}
\end{eqnarray}
where $g$ is the HLS gauge coupling, and
${\mathcal L}_\mu$ and ${\mathcal R}_\mu$ denote the external gauge fields
gauging the $G_{\rm global}$ symmetry.

The HLS Lagrangian at the leading order
is given by~\cite{BKUYY-BKY:88}
\begin{equation}
{\mathcal L} = F_\pi^2 \, \mbox{tr} 
\left[ \hat{\alpha}_{\perp\mu} \hat{\alpha}_{\perp}^\mu \right]
+ F_\sigma^2 \, \mbox{tr}
\left[ 
  \hat{\alpha}_{\parallel\mu} \hat{\alpha}_{\parallel}^\mu
\right]
+ {\mathcal L}_{\rm kin}(\rho_\mu) \ ,
\label{Lagrangian}
\end{equation}
where ${\mathcal L}_{\rm kin}(\rho_\mu)$ denotes the kinetic term of
$\rho_\mu$ 
and
\begin{eqnarray}
&&
\hat{\alpha}_{\perp,\parallel}^\mu =
( D_\mu \xi_{\rm R} \cdot \xi_{\rm R}^\dag \mp 
  D_\mu \xi_{\rm L} \cdot \xi_{\rm L}^\dag
) / (2i)
\ .
\end{eqnarray}

\subsection{Renormalization Group Equations}
\label{ssec:RGE}

At one-loop level
the Lagrangian (\ref{Lagrangian}) 
generates the ${\mathcal O}(p^4)$
contributions including the divergent contributions which
are renormalized by three leading order
parameters $F_\pi$, $a$ and $g$ (and parameters of 
${\mathcal O}(p^4)$ Lagrangian).
As was stressed in Ref.~\refcite{HY:PRep},
it is important to include effects of quadratic divergences
into the resultant RGEs
for studying the phase structure.
As is well known, the naive momentum cutoff violates the chiral
symmetry.
Then, it is important to use a way to
include quadratic divergences 
consistently with the chiral symmetry.
By adopting the dimensional regularization and identifying 
the quadratic
divergences with the presence of poles of ultraviolet origin at
$n=2$~\cite{Veltman}, the RGEs for three leading order parameters
are expressed as~\cite{HY:conformal,HY:WM,HY:PRep}
\begin{eqnarray}
{\mathcal M} \frac{dF_\pi^2}{d{\mathcal M}} &=& 
C\left[3a^2g^2F_\pi^2 +2(2-a){\mathcal M}^2 \right] 
\ ,
\nonumber\\
{\mathcal M} \frac{da}{d{\mathcal M}} &=&-C
(a-1) \left [3a(1+a)g^2-(3a-1)\frac{{\mathcal M}^2}{F_\pi^2} \right]
\ ,
\nonumber\\
{\mathcal M}\frac{d g^2}{d {\mathcal M}}&=& -C\frac{87-a^2}{6}g^4
\ ,
\label{rge}
\end{eqnarray}
where $C = N_f/\left[2(4\pi)^2\right]$ and ${\mathcal M}$ is the
renormalization point.~\footnote{%
  In Refs.~\refcite{HY:conformal,HY:WM,HY:PRep}, the renormalization
  point is expressed by $\mu$.  In this write-up, however,
  I preserve $\mu$ for expressing the chemical potential.
}
It should be noted that the point $(g\,,\,a)=(0,1)$
is the fixed point of 
the RGEs in Eq.~(\ref{rge})
which plays an essential role to realize the VM in the following
analysis of the chiral restoration.

\section{Wilsonian Matching}
\label{sec:WM}

In Ref.~\refcite{HY:WM},
the Wilsonian matching was proposed 
to
determine the bare parameters of the HLS Lagrangian
by matching the HLS to
the underlying QCD.
In this section, I briefly summarize 
the Wilsonian matching.

The Wilsonian matching proposed in Ref.~\refcite{HY:WM} 
is done by matching
the axialvector and vector current correlators derived from the
HLS with those by the operator product expansion (OPE) in
QCD at the matching scale $\Lambda$.~\footnote{%
  For the validity of the expansion in the HLS, the
  matching scale $\Lambda$ must be smaller than the chiral symmetry
  breaking scale $\Lambda_\chi$.
}
The axialvector and vector current correlators in the OPE 
up until ${\mathcal O}(1/Q^6)$
are expressed as~\cite{SVZ}
\begin{eqnarray}
&&
\Pi_A^{\rm(QCD)}(Q^2) = \frac{1}{8\pi^2}
\left( \frac{N_c}{3} \right)
\Biggl[
  - \left( 
      1 +  \frac{3(N_c^2-1)}{8N_c}\, \frac{\alpha_s}{\pi} 
  \right) \ln \frac{Q^2}{\mu^2}
\nonumber\\
&& \quad
  {}+ \frac{\pi^2}{N_c} 
    \frac{
      \left\langle 
        \frac{\alpha_s}{\pi} G_{\mu\nu} G^{\mu\nu}
      \right\rangle
    }{ Q^4 }
  {}+ \frac{\pi^3}{N_c} \frac{96(N_c^2-1)}{N_c^2}
    \left( \frac{1}{2} + \frac{1}{3N_c} \right)
    \frac{\alpha_s \left\langle \bar{q} q \right\rangle^2}{Q^6}
\Biggr]
\ ,
\label{Pi A OPE}
\\
&&
\Pi_V^{\rm(QCD)}(Q^2) = \frac{1}{8\pi^2}
\left( \frac{N_c}{3} \right)
\Biggl[
  - \left( 
      1 +  \frac{3(N_c^2-1)}{8N_c}\, \frac{\alpha_s}{\pi} 
  \right) \ln \frac{Q^2}{\mu^2}
\nonumber\\
&& \quad
  {}+ \frac{\pi^2}{N_c} 
    \frac{
      \left\langle 
        \frac{\alpha_s}{\pi} G_{\mu\nu} G^{\mu\nu}
      \right\rangle
    }{ Q^4 }
  {}- \frac{\pi^3}{N_c} \frac{96(N_c^2-1)}{N_c^2}
    \left( \frac{1}{2} - \frac{1}{3N_c} \right)
    \frac{\alpha_s \left\langle \bar{q} q \right\rangle^2}{Q^6}
\Biggr]
\ ,
\label{Pi V OPE}
\end{eqnarray}
where $\mu$ is the renormalization scale of QCD
and we
wrote the $N_c$-dependences explicitly
(see, e.g., Ref.~\refcite{Bardeen-Zakharov}).
In the HLS the same correlators are 
well described by the tree contributions with including
${\mathcal O}(p^4)$ terms 
when the momentum is around the matching scale, $Q^2 \sim \Lambda^2$:
\begin{eqnarray}
\Pi_A^{\rm(HLS)}(Q^2) &=&
\frac{F_\pi^2(\Lambda)}{Q^2} - 2 z_2(\Lambda)
\ ,
\label{Pi A HLS}
\\
\Pi_V^{\rm(HLS)}(Q^2) &=&
\frac{
  F_\sigma^2(\Lambda)
}{
  M_\rho^2(\Lambda) + Q^2
} 
\left[ 1 - 2 g^2(\Lambda) z_3(\Lambda) \right]
- 2 z_1(\Lambda)
\ ,
\label{Pi V HLS}
\end{eqnarray}
where the bare $\rho$ mass $M_\rho(\Lambda)$ is defined as
\begin{equation}
M_\rho^2(\Lambda) \equiv g^2(\Lambda) F_\sigma^2(\Lambda)
\ .
\label{on-shell cond 5}
\end{equation}

I require that current correlators in the HLS
in Eqs.~(\ref{Pi A HLS}) and (\ref{Pi V HLS})
can be matched with those in QCD in 
Eqs.~(\ref{Pi A OPE}) and (\ref{Pi V OPE}).
Of course,
this matching cannot be made for any value of $Q^2$, 
since the $Q^2$-dependences of the current correlators 
in the HLS are completely
different from those in the OPE:
In the HLS the derivative expansion (in {\it positive} power of $Q$)
is used, and the expressions for
the current correlators are valid in the low energy region.
The OPE, on the other hand, is an asymptotic expansion
(in {\it negative} power of $Q$), and it is
valid in the high energy region.
Since I calculate the current correlators in the HLS including the
first non-leading order [${\mathcal O}(p^4)$], 
I expect that I can match
the correlators with those in the OPE up until the first derivative.
Then I obtain the following Wilsonian matching 
conditions~\cite{HY:WM,HY:PRep}
\begin{eqnarray}
&&
   \frac{F^2_\pi (\Lambda)}{{\Lambda}^2} 
  = \frac{1}{8{\pi}^2} \left( \frac{N_c}{3} \right) 
  \Biggl[
    1 +
    \frac{3(N_c^2-1)}{8N_c}\, \frac{\alpha_s}{\pi} 
    + \frac{2\pi^2}{N_c} 
      \frac{
        \left\langle 
          \frac{\alpha_s}{\pi} G_{\mu\nu} G^{\mu\nu}
        \right\rangle
      }{ \Lambda^4 }
\nonumber\\
&& \qquad\qquad\qquad\qquad
    {}+ \frac{288\pi(N_c^2-1)}{N_c^3}
      \left( \frac{1}{2} + \frac{1}{3N_c} \right)
      \frac{\alpha_s \left\langle \bar{q} q \right\rangle^2}
           {\Lambda^6}
  \Biggr]
\ ,
\label{match A}
\\
&&
   \frac{F^2_\sigma (\Lambda)}{{\Lambda}^2}
        \frac{{\Lambda}^4[1 - 2g^2(\Lambda)z_3(\Lambda)]}
         {({M_\rho}^2(\Lambda) + {\Lambda}^2)^2}
\nonumber\\
&& \qquad
= \frac{1}{8\pi^2} \left( \frac{N_c}{3} \right)
  \Biggl[
    1 +
    \frac{3(N_c^2-1)}{8N_c}\, \frac{\alpha_s}{\pi} 
    + \frac{2\pi^2}{N_c} 
      \frac{
        \left\langle 
          \frac{\alpha_s}{\pi} G_{\mu\nu} G^{\mu\nu}
        \right\rangle
      }{ \Lambda^4 }
\nonumber\\
&& \qquad\qquad\qquad\qquad
    {}- \frac{288\pi(N_c^2-1)}{N_c^3}
      \left( \frac{1}{2} - \frac{1}{3N_c} \right)
      \frac{\alpha_s \left\langle \bar{q} q \right\rangle^2}
           {\Lambda^6}
  \Biggr]
\ ,
\label{match V}
\\
&&
   \frac{F^2_\pi (\Lambda)}{{\Lambda}^2} - 
            \frac{F^2_\sigma (\Lambda)[1 -
              2g^2(\Lambda)z_3(\Lambda)]}
             {{M_\rho}^2(\Lambda) + {\Lambda}^2} -
            2[z_2(\Lambda) - z_1(\Lambda)] 
\nonumber\\
&& \qquad
  =  \frac{4\pi(N_c^2-1)}{N_c^2}
  \frac{\alpha_s \left\langle \bar{q} q \right\rangle^2}{\Lambda^6}
\ .
\label{match z}
\end{eqnarray}
The above three equations (\ref{match A}), (\ref{match V}) and 
(\ref{match z}) are the Wilsonian matching conditions
proposed in Ref.~\refcite{HY:WM}.
These determine several bare parameters of the HLS without 
much ambiguity.
Especially, the first condition (\ref{match A}) 
determines the value of the bare $\pi$ decay constant
$F_\pi(\Lambda)$ directly from QCD.

Once the bare parameters are determined through the Wilsonian
matching, one can calculate several physical quantities
for $\pi$ and $\rho$
using the Wilsonian RGEs summarized in 
subsection~\ref{ssec:RGE},
in excellent agreement with the experiments for real-life
QCD with $N_f=3$
(see, for details, Refs.~\refcite{HY:WM,HY:PRep}).
This encourages
us to perform the analysis for other situations such as 
larger $N_f$ and finite temperature and/or density 
up to near the critical point, based on the Wilsonian matching.

\section{Vector Manifestation in Large $N_f$ QCD}
\label{sec:VMNf}

In this section, I briefly summarize how the vector manifestation
(VM) is realized near the critical point 
in the large $N_f$ QCD.

The chiral symmetry restoration in Wigner realization should
be characterized by 
\begin{equation}
F_\pi(0) =0
\label{zerofpi}
\end{equation}
and the equality of the vector and axialvector current
correlators in the underlying QCD:
\begin{equation}
\Pi_V(Q^2) = \Pi_A(Q^2) \ ,
\end{equation}
which is in accord with $\langle \bar{q} q \rangle = 0$ in 
Eqs.~(\ref{Pi A OPE}) and (\ref{Pi V OPE}).
On the other hand, the same current correlators are described 
in terms of the HLS model for energy lower than the cutoff $\Lambda$: 
When we approach to the critical point {\it from the broken phase (NG
phase)},  the axialvector current correlator 
is still dominated by 
the massless $\pi$ as the NG boson, 
while the vector current correlator is
by the massive $\rho$. 
In such a case, there exists a scale $\Lambda$ around which the
current correlators are well described by the forms given in
Eqs.~(\ref{Pi A HLS}) and (\ref{Pi V HLS}).
Then, through
the Wilsonian matching discussed in section~\ref{sec:WM},
the bare parameters of the HLS are determined.
At the critical point
the quark condensate vanishes,
$\left\langle \bar{q} q \right\rangle \rightarrow 0$, while
the gluonic condensate 
$\left\langle 
\frac{\alpha_s}{\pi} G_{\mu\nu} G^{\mu\nu}
\right\rangle$
is independent of the renormalization point of
QCD and hence  it is expected not to vanish.
Then Eq.~(\ref{match A}) reads
\begin{eqnarray}
&&
F_\pi^2(\Lambda)
\rightarrow (F_\pi^{\rm crit})^2\equiv   
\frac{N_c}{3}\left(\frac{\Lambda}{4\pi}\right)^2
\cdot 2
(  1 + \delta_A^{\rm crit})\ \ne 0 
\nonumber
,\\
&&\delta_A^{\rm crit}
\equiv 
\frac{3(N_c^2-1)}{8N_c} \,\frac{\alpha_s}{\pi}
  + \frac{2\pi^2}{N_c} 
    \frac{
      \left\langle 
        \frac{\alpha_s}{\pi} G_{\mu\nu} G^{\mu\nu}
      \right\rangle
    }{ \Lambda^4 }
>0\quad (\ll 1)
\ ,
\label{match A VM}
\end{eqnarray}
implying that {\it matching with QCD dictates}
\begin{equation}
F_\pi^2(\Lambda) \ne 0
\label{nvf_pi}
\end{equation}
{\it even at the critical point}~\cite{HY:VM} where
$F_\pi^2(0)=0$.
I should note that
{\it $F_\pi(\Lambda)$ is not an order parameter but just a
parameter of the bare HLS Lagrangian} defined at the cutoff
$\Lambda$ where the
matching with QCD is made,
while $F_\pi(0)$ is the order parameter.

Let me obtain further 
constraints on other bare parameters of the HLS   
through the Wilsonian matching for the currents correlators.
The constraints on other parameters defined at $\Lambda$
come from the fact that $\Pi_A^{\rm (QCD)}$ 
and $\Pi_V^{\rm (QCD)}$ in Eqs.~(\ref{Pi A OPE}) and (\ref{Pi V OPE})
agree with each other for any value of $Q^2$ when the chiral symmetry
is restored with $\left\langle \bar{q} q \right\rangle \rightarrow 0$.
Thus, I require that
$\Pi_A^{\rm (HLS)}$ and $\Pi_V^{\rm (HLS)}$ in Eqs.~(\ref{Pi A HLS})
and (\ref{Pi V HLS})
agree with each other for {\it any value of $Q^2$}
(near $\Lambda^2$).
Under the condition~(\ref{nvf_pi}),
this agreement is satisfied only if the following conditions are met:
\begin{eqnarray}
&& g(\Lambda) \rightarrow 0 \ , 
\label{vector condition:g}
\\
&& a(\Lambda) = \frac{ F_\sigma^2(\Lambda) }{ F_\pi^2(\Lambda) }
\rightarrow 1 \ , 
\label{vector condition:a}
\\
&& z_1(\Lambda) - z_2(\Lambda) \rightarrow 0 \ ,
\label{vector condition:z}
\\
&& 
F_\pi^2(\Lambda) \rightarrow (F_\pi^{\rm crit})^2
=
\frac{N_c}{3}
\left(\frac{\Lambda}{4\pi}\right)^2\cdot 2(1+\delta_A^{\rm crit}) 
\neq 0
\ .
\label{vector condition:fp}
\end{eqnarray}
These conditions, may be called ``VM conditions'',
follow solely from the requirement of the equality of
the vector and axialvector currents correlators (and the Wilsonian matching)
without explicit requirement of Eq.~(\ref{zerofpi}), 
and are actually a precise expression of the
VM in terms of the {\it bare}
HLS parameters for the Wigner realization in 
QCD.~\cite{HY:VM}

Let me show that the above VM conditions actually leads to
the chiral restoration when $N_f$ approaches the critical number.
Since the values in 
Eqs.~(\ref{vector condition:g})--(\ref{vector condition:z}) 
coincide with 
those at the fixed points of the RGEs~\cite{HY:VM,HY:PRep},
the RGE for $F_\pi^2$ in Eq.~(\ref{rge}) becomes a simple form which
is readily solved as
\begin{eqnarray}
&&F_\pi^2(0)
= 
F_\pi^2(\Lambda) - 
\frac{N_f\Lambda^2}{2(4\pi)^2} \rightarrow 
(F_\pi^{\rm crit})^2
- \frac{N_f\Lambda^2}{2(4\pi)^2} 
\ .
\label{Fpirun}
\end{eqnarray}
The $(F_\pi^{\rm crit})^2$ in the above relation,
which is determined through the 
Wilsonian matching as in Eq.~(\ref{vector condition:fp}),
is almost independent of $N_c$ as well as $N_f$~\cite{HY:VM,HY:PRep}.
The second term, on the other hand, linearly increase with
$N_f$.
This implies that
one can have 
\begin{equation}
F_\pi^2(0)\rightarrow 0 \ ,
\label{fp0 0}
\end{equation}
for $N_f \rightarrow N_f^{\rm crit}$.
Then the chiral restoration $F_\pi^2(0) \rightarrow 0$ is actually
{\it derived within the dynamics 
of the HLS model itself} solely from  the requirement 
of the Wilsonian matching.

I estimate the number of critical flavor, $N_f^{\rm crit}$.
By combining Eqs.~(\ref{Fpirun}) and (\ref{fp0 0})
with the VM condition~(\ref{vector condition:fp}),
$N_f^{\rm crit}$ is expressed in terms of the parameters in the
OPE as
\begin{equation}
N_f^{\rm crit} = 2 (4\pi)^2\frac{(F_\pi^{\rm crit})^2}{\Lambda^2}
=\frac{N_c}{3} \cdot 4 (1+\delta_A^{\rm crit}) \ ,
\label{Nfcrit0}
\end{equation}
where the form of $\delta_A^{\rm crit}$ is given in 
Eq.~(\ref{match A VM}).
By using 
$\frac{3}{N_c} \langle \frac{\alpha_s}{\pi} 
  G_{\mu\nu} G^{\mu\nu} \rangle = 
  0.012\,\mbox{GeV}^4$~\cite{SVZ,Bardeen-Zakharov}
and $\alpha_s = 0.56$ for 
$(\Lambda,\Lambda_{\rm QCD}) = (1.1,0.4)\,\mbox{GeV}$
as a typical example,
the number of critical flavor is estimated 
as~\cite{HY:VM,HY:PRep}
\begin{equation}
N_f^{\rm crit} \simeq 5 \, \left( \frac{N_c}{3} \right) \ .
\end{equation}

Now, I study the vector meson mass and decay constant
near the critical point.
As I discussed above,
the values in 
Eqs.~(\ref{vector condition:g})--(\ref{vector condition:z}) 
coincide with 
those at the fixed points of the RGEs~\cite{HY:VM,HY:PRep},
so that
the parameters remains the same for any scale,
and hence even at $\rho$ on-shell point:
\begin{eqnarray}
g(m_\rho) \rightarrow 0 \ , 
\quad
a(m_\rho) \rightarrow 1 \ , 
\quad
z_1(m_\rho) - z_2(m_\rho) \rightarrow 0 \ ,
\label{vector condition:on-shell}
\end{eqnarray}
where $m_\rho$ is determined from the on-shell condition:
\begin{equation}
m_\rho^2 = a(m_\rho) g^2(m_\rho) F_\pi^2(m_\rho) \ .
\label{on-shell condition 6}
\end{equation}
Then, the first condition in 
Eq.~(\ref{vector condition:on-shell})
together with the above on-shell condition immediately leads to
\begin{eqnarray}
&&
m_\rho^2 \rightarrow 0\ .
\label{VM mrho}
\end{eqnarray}
The second condition in 
Eq.~(\ref{vector condition:on-shell}) is rewritten as
$F_\sigma^2(m_\rho) /F_\pi^2(m_\rho) \rightarrow 1$, and 
Eq.~(\ref{VM mrho}) implies $F_\pi^2(m_\rho) \rightarrow F_\pi^2(0)$.
Thus, 
\begin{eqnarray}
&&
F_\sigma^2(m_\rho) /F_\pi^2(0) \rightarrow 1 \ ,
\label{VM fsig}
\end{eqnarray}
namely, the pole residues of $\pi$ and $\rho$ become identical.
Then the VM defined by Eq.~(\ref{VM def}) does follow.
Note that I have used only the requirement of Wigner realization in
QCD 
through the Wilsonian matching 
and {\it arrived uniquely at VM but not GL manifestation 
\'a la linear 
sigma model}.
The crucial ingredient to exclude the GL manifestation as a chiral
restoration in QCD was the Wilsonian 
matching, particularly Eq.~(\ref{nvf_pi}).

\section{Vector Manifestation in Hot Matter}
\label{sec:VMT}

In this section,
I briefly summarize
how the vector manifestation (VM) is realized in
hot matter following Ref.~\refcite{HS:VMT}.
(Details of the calculation can be seen in 
Refs.~\refcite{Harada-Kim-Rho-Sasaki,Harada-Sasaki:2}.)

For the VM in large $N_f$ QCD, 
the VM conditions derived from the Wilsonian matching
play important roles as I explained in the previous section.
In Ref.~\refcite{HS:VMT},
the Wilsonian matching was extended to the hot matter
calculation to determine the bare parameters of the HLS Lagrangian.
It was stressed that the matching procedure leads to the
{\it intrinsic temperature dependences} of the bare parameters.
Especially,
the intrinsic temperature dependence of the bare $\pi$ decay
constant is determined by putting the possible temperature
dependences on the gluon and quark condensate
in Eq.~(\ref{match A})~\cite{HS:VMT}:
\begin{eqnarray}
\!\!\!\!
   \frac{F^2_\pi (\Lambda ;T)}{{\Lambda}^2} 
    = \frac{1}{8{\pi}^2} 
    \Biggl[
        1 +  \frac{\alpha_s}{\pi} 
        + \frac{2\pi^2}{3} 
           \frac{\langle \frac{\alpha _s}{\pi}
            G_{\mu \nu}G^{\mu \nu} \rangle_T }
                                {{\Lambda}^4} 
       {}+ \pi^3 \,\frac{1408}{27}
                                    \frac{\alpha _s{\langle \bar{q}q
                                         \rangle }^2_T}
                                     {{\Lambda}^6}
                                    \Biggr]
\ , 
\label{eq:WMC A}
\end{eqnarray}
where I took $N_c=3$.

Now, let me consider the Wilsonian matching near the
critical temperature $T_c$
with assuming that the quark condensate becomes zero
continuously for $T\rightarrow T_c$.
First, note that
the Wilsonian matching condition~(\ref{eq:WMC A}) 
provides
\begin{equation}
  \frac{F^2_\pi (\Lambda ;T_c)}{{\Lambda}^2} 
  = \frac{1}{8{\pi}^2}
  \left[
    1 + \frac{\alpha _s}{\pi}
    + \frac{2{\pi}^2}{3}
      \frac{\langle \frac{\alpha _s}{\pi} 
            G_{\mu \nu}G^{\mu \nu} \rangle_{T_c} }
      {{\Lambda}^4}
  \right]
 \neq 0 
\ ,
\label{eq:WMC A Tc}
\end{equation}
even at the critical point where the on-shell $\pi$
decay constant vanishes 
by adding the quantum corrections through
the RGE including the quadratic divergence~\cite{HY:conformal}
and hadronic thermal-loop
corrections~\cite{HS:VMT}.
Second, I note that the axialvector and vector current correlators
$\Pi_A^{(\rm{QCD})}$ and $\Pi_V^{(\rm{QCD})}$
derived by the OPE
agree with each other for any value of $Q^2$.
Thus I require that
these current correlators in the HLS are
equal at the critical point
for any value of $Q^2\ \mbox{around}\ {\Lambda}^2$.
This requirement 
$\Pi_A^{(\rm{HLS})}=\Pi_V^{(\rm{HLS})}$ is satisfied
if the following conditions are met~\cite{HS:VMT}: 
\begin{eqnarray}
&&
g(\Lambda;T) \mathop{\longrightarrow}_{T \rightarrow T_c} 0 \ ,
\qquad
a(\Lambda;T) \mathop{\longrightarrow}_{T \rightarrow T_c} 1 \ ,
\nonumber\\
&&
z_1(\Lambda;T) - z_2(\Lambda;T) 
\mathop{\longrightarrow}_{T \rightarrow T_c} 0 \ .
\label{g a z12:VMT}
\end{eqnarray}
These conditions (``VM conditions in hot matter'')
for the bare parameters
are converted into the
conditions for the on-shell parameters through the Wilsonian RGEs.
Since $g=0$ and $a=1$ are separately the fixed points of the RGEs for
$g$ and $a$~\cite{HY:conformal},
the on-shell parameters also satisfy
$(g,a)=(0,1)$, and thus the parametric $\rho$ mass
satisfies $M_\rho = 0$.

Now, let me
include the hadronic thermal effects to obtain the $\rho$ pole
mass near the critical temperature.
As I explained above,
the intrinsic temperature dependences imply that
$M_\rho/T \rightarrow 0$
for $T \rightarrow T_c$,
so that the $\rho$ pole mass near the
critical temperature is expressed as~\cite{HS:VMT}
\begin{eqnarray}
&& m_\rho^2(T)
  = M_\rho^2 +
  g^2 N_f \, \frac{15 - a^2}{144} \,T^2
\ .
\label{mrho at T 2}
\end{eqnarray}
Since $a \simeq 1$ near the restoration point,
the second term is positive. 
Then the $\rho$ pole mass $m_\rho$
is bigger than the parametric
$M_\rho$ due to the hadronic thermal corrections.
Nevertheless, 
{\it the intrinsic temperature dependence determined by the
Wilsonian matching requires
that the $\rho$ becomes massless at the
critical temperature}:
\begin{eqnarray}
&&
m_\rho^2(T)
\rightarrow 0 \ \ \mbox{for} \ T \rightarrow T_c \ ,
\end{eqnarray}
since the first term in Eq.~(\ref{mrho at T 2})
vanishes as $M_\rho\rightarrow 0$, and the second
term also vanishes since $g\rightarrow 0$ for $T \rightarrow T_c$.
This implies that
{\it the vector manifestation (VM) actually
occurs at the critical
temperature}~\cite{HS:VMT}.

\section{Vector Manifestation in Dense Matter}
\label{sec:VMm}

In this section, I briefly summarize how the 
VM is realized in dense QCD following Ref.~\refcite{HKR}.

In Ref.~\refcite{HKR}, 
following the picture shown in Ref.~\refcite{Brown-Rho:01b},
the quasiquark degree of freedom is added into the HLS Lagrangian
near the critical chemical potential
with assuming that its mass $m_q$ becomes small 
($m_q \rightarrow 0$).
The Lagrangian introduced in Ref.~\refcite{HKR} for
including one quasiquark field $\psi$
and one anti-quasiquark field $\bar{\psi}$ is
counted as ${\mathcal O}(p)$ and given by
\begin{eqnarray}
 {\mathcal L}_{Q} 
= 
 \bar \psi( iD_\mu \gamma^\mu
       + \mu \gamma^0 -m_q )\psi
  {}+ \bar \psi \left(
  \kappa\gamma^\mu \hat{\alpha}_{\parallel \mu}
+ \lambda\gamma_5\gamma^\mu \hat{\alpha}_{\perp\mu} \right)
        \psi 
\ ,
\label{lagbaryon}
 \end{eqnarray}
where $\mu$ is the chemical potential,
$D_\mu\psi=(\partial_\mu -i g \rho_\mu)\psi$ and $\kappa$ and
$\lambda$ are constants to be specified later.

Inclusion of the quasiquark changes the RGEs for $F_\pi$, $a$ 
and $g$.
Furtheremore, the RGE for the quasiquark mass $m_q$ should be
considered simultaneously.
The explicit forms of the RGEs are shown in Eq.~(7)
of Ref.~\refcite{HKR},
which show 
that,
although $g=0$ and $a=1$ are not separately
the fixed points of the RGEs for $g$ and $a$, 
$(g,a,m_q) = (0,1,0)$ is a fixed point of the coupled RGEs for 
$g$, $a$ and $m_q$.

Let me consider the 
{\it intrinsic density dependences} of the 
bare parameters of the HLS Lagrangian.
Similarly to the intrinsic temperature dependences in hot QCD,
the intrinsic density dependences are 
introduced through the Wilsonian matching.
Noting that the quasiquark does not contribute to the current
correlators at {\it bare level}, one arrives at the following
``VM conditions in dense matter'' similar to the one
in hot matter in Eq.~(\ref{g a z12:VMT}) near the critical 
chemical potential $\mu_c$~\cite{HKR}:
\begin{eqnarray}
&&
g(\Lambda;\mu) \mathop{\longrightarrow}_{\mu \rightarrow \mu_c} 0 \ ,
\qquad
a(\Lambda;\mu) \mathop{\longrightarrow}_{\mu \rightarrow \mu_c} 1 \ ,
\nonumber\\
&&
z_1(\Lambda;\mu) - z_2(\Lambda;\mu) 
\mathop{\longrightarrow}_{\mu \rightarrow \mu_c} 0 \ .
\label{g a z12:VMm}
\end{eqnarray}
These conditions are converted into the conditions for the
on-shell parameters throught the RGEs.
Since 
$(g,a,m_q) = (0,1,0)$ is a fixed point of the coupled RGEs for 
$g$, $a$ and $m_q$, the above conditions together with the assumption
$m_q \rightarrow 0$ for $\mu \rightarrow \mu_c$ imply that
the on-shell parameters behave as
$g\rightarrow 0$ and $a\rightarrow 1$, and thus the parametric
$\rho$ mass vanishes for $\mu \rightarrow \mu_c$: 
$M_\rho \rightarrow 0$.

Now, let me study the $\rho$ pole mass near $\mu_c$
by including the hadronic dense-loop correction from the
quasiquark.
By taking $(g,a,m_q) \rightarrow (0,1,0)$ 
in the quasiquark loop contribution, 
the $\rho$ pole mass is expressed as
\begin{equation}
m_\rho^2(\mu) = M_\rho^2(\mu) + g^2 \,
\frac{(1-\kappa)^2}{6\pi^2} \,\mu^2 \ .
\end{equation}
Since $M_\rho(\mu) \rightarrow 0$ and $g\rightarrow0$ for
$\mu \rightarrow \mu_c$ due to the intrinsic density dependence,
the above expression implies that 
the $\rho$ pole mass vanishes at the critical chemical potential,
i.e.,
the VM is realized in dense matter:
\begin{equation}
m_\rho(\mu) 
\rightarrow 0 \ \ \mbox{for} \ \mu \rightarrow \mu_c \ .
\end{equation}

\section{Summary}
\label{sec:sum}

In this write-up,
I first summarized main features of the VM 
which was recently proposed
as a novel manifestation of the Wigner realization of 
chiral symmetry in which the symmetry is restored at the critical
point by the massless degenerate pion (and its flavor partners) and
the $\rho$ meson (and its flavor partners) as the chiral partner.
Then, 
I have shown how the VM is formulated in 
the effective field theory of QCD based on the HLS
and
realized in the large $N_f$ QCD as well as
in hot and/or dense QCD.

Finally, I would like to note that
a detailed
review of loop expansion in the HLS and the VM is given
in Ref.~\refcite{HY:PRep}, and that
several predictions of the VM in hot QCD are shown in
Refs.~\refcite{Harada-Kim-Rho-Sasaki} and
\refcite{Harada-Sasaki:2} as well as in the 
write-up by Prof.~Rho~\cite{Rho:SCGT02}
 in the proceedings of SCGT02.

\section*{Acknowledgement}

I would like to thank 
Dr. Youngman Kim,
Prof. Mannque Rho,
Dr. Chihiro Sasaki and
Prof. Koichi Yamawaki for collaborations on the works
on which this talk is based.

\end{document}